# A 750 mW, continuous-wave, solid-state laser source at 313 nm for cooling and manipulating trapped $^9$Be$^+$ ions

A. C. Wilson[1], C. Ospelkaus[1,2], A. P. VanDevender[1,3], J. A. Mlynek[1,4], K. R. Brown[1], D. Leibfried[1] and D. J. Wineland[1]

[1]*National Institute of Standards & Technology, Time & Frequency Division, 325 Broadway, Boulder CO 80305, USA*

[2]*Present address: The Institute of Quantum Optics, Leibniz Universität Hannover, Welfengarten 1, D-30167 Hannover, Germany, and Physikalisch-Technische Bundesanstalt, Braunschweig Bundesallee 100, D-38116 Braunschweig, Germany*

[3] *Present address: A. P. VanDevender - sig@netdot.net. Present address: Halcyon Molecular, 505 Penobscot Drive, Redwood City, CA 94063, USA*

[4] *Present address: J. A. Mlynek – mlynek@phys.ethz.ch. Present address: Department of Physics, Laboratory for Solid State Physics, ETH, Zurich, Switzerland*

**Abstract**
We present a solid-state laser system that generates 750 mW of continuous-wave single-frequency output at 313 nm. Sum-frequency generation with fiber lasers at 1550 nm and 1051 nm produces up to 2 W at 626 nm. This visible light is then converted to UV by cavity-enhanced second-harmonic generation. The laser output can be tuned over a 495 GHz range, which includes the $^9$Be$^+$ laser cooling and repumping transitions. This is the first report of a narrow-linewidth laser system with sufficient power to perform fault-tolerant quantum-gate operations with trapped $^9$Be$^+$ ions by use of stimulated Raman transitions.

## 1    Introduction

Two of the primary objectives in quantum information processing and computing, are scaling to large numbers of quantum gates, and achieving fault-tolerant gate operation [1-2]. A promising approach to achieving both objectives is to use trapped ions, in which the quantum information is encoded on internal atomic states [3]. Efforts to improve ion-trap scalability have focused on multi-zone arrays [4-5], with complex surface-electrode geometries that provide multiple trapping zones [6]. These traps include control electrodes that enable the shuttling of ions between zones that are used to perform gate operations, state detection, and information storage. For fault-tolerant two-qubit (quantum bit) gate operations, error-correction protocols have been proposed [7-8], but these require a sufficiently low error per gate (typically assumed to be less that $10^{-4}$), and this threshold has not yet been achieved. In one approach, qubits are encoded into ground-state hyperfine states, since these are very well isolated from environmental effects that cause memory error. However, gate operations are usually performed via optical transitions with laser beams, leading to spontaneous emission that dominates gate error. Spontaneous emission is reduced by a large detuning from atomic resonance, but then higher laser powers required to maintain the gate speed. For example, Ozeri *et al*. calculate that with the commonly trapped ion species of $^9$Be$^+$, $^{25}$Mg$^+$, and $^{43}$Ca$^+$, in order to reach the fault-tolerant regime for a two-qubit phase gate, one needs narrow-linewidth, continuous-wave (cw) laser power in the range of 140 mW to 540 mW (and detuning from atomic resonance on the THz scale) [9].

The most challenging aspect of developing laser sources for this purpose is that most of the wavelengths are in the UV region. For trapped $^{43}$Ca$^+$ ions, the required 729 nm or 397 nm light can be generated directly with semiconductor lasers; but for the shorter wavelengths needed for most other trapped-ion species, the traditional approach has been to frequency-double the visible output from a ring dye laser. While solid-state lasers have replaced gas and dye lasers in many spectral regions, some wavelengths have remained difficult to produce. In 2006, Friedenauer *et al*. described a high-power, solid-state laser system for generating the 280 nm light needed to laser-cool and detect trapped $^{25}$Mg$^+$ ions [10]. In their scheme, the output from a fiber laser at 1120 nm is frequency-doubled twice to produce 275 mW at 280 nm. For $^9$Be$^+$ ions, light at 313 nm is required, and since high-power, narrow-band fiber lasers are not available at 1252 nm, it is necessary to adopt an approach different from that used by Friedenauer *et al*.

In 2002, Schnitzler *et al*. demonstrated a solid-state laser system producing 33 mW at 313 nm [11]. Their setup includes a frequency-doubled Nd:YAG laser at 532 nm, a titanium sapphire laser at 760 nm, and sum-frequency generation (SFG) in an enhancement cavity resonant at both the pump (532 nm) and signal (760 nm) wavelengths. More recently, Vasilyev *et al*. described a solid-state-laser-based source generating 100 mW at 313 nm [12]. In their scheme, a high-power fiber laser system at 1565 nm is followed by two stages of second-harmonic generation (SHG) and two stages of SFG, producing the fifth harmonic of 1565 nm at 313 nm.

In this paper we describe a solid-state laser system that generates the high UV power necessary for fault-tolerant quantum gates with trapped $^9$Be$^+$ ions. The setup can also be used for laser cooling and precision spectroscopy with beryllium ions. Our approach is to generate light at 626 nm by use of SFG and then frequency-double to 313 nm. The traditional workhorse for generating 626 nm in our laboratory is the frequency-stabilized ring dye laser. This produces up to ~1 W at 626 nm, and after fiber optic delivery to our frequency doubling setup, we typically obtain up to ~150 mW at 313 nm. Recently, we have developed a solid-state alternative to the dye laser that uses SFG with the output of two narrow-linewidth, high-power fiber lasers operating at the relatively standard wavelengths of 1550 nm and 1051 nm. The visible light is then frequency-doubled to 313 nm. The laser system produces twice as much 626 nm power as our dye lasers, and more than five times the UV power that Ozeri *et al*. predict is necessary for performing fault-tolerant two-qubit phase gates with $^9$Be$^+$. This power overhead compensates for losses in an optical control system and is sufficient to perform two-qubit gate operations.

## 2    Sum-frequency generation of 626 nm light

Our approach to generating 626 nm is similar to the SFG scheme reported by Hart *et al*. [13], except that here the pump and signal are generated by two separate, narrow-bandwidth, near-infrared (NIR) fiber lasers, so that the two input beam shapes can be adjusted individually for maximum SFG output. A schematic diagram of the optical setup is shown in Fig. 1. A Koheras KOH1895 Boostik fiber laser[1] produces up to 4.90 W at 1051 nm, with a specified linewidth of < 70 kHz. Coarse (slow) tuning of this laser is performed with a temperature adjustment (coefficient 5.4 GHz/K, range 108.6 GHz). Fine (fast) tuning can be performed with a piezo-electric transducer (coefficient 23 MHz/V, range 4.6 GHz). For the measurements reported here, the laser was tuned to 1051.140 nm (vacuum wavelength). At maximum output, the ratio of signal to amplified spontaneous emission (ASE) is specified to be > 13 dB. The laser's output optical fiber has an integrated collimator unit that produces a Gaussian beam size (radius) equal to $1.05 \pm 0.02$ mm.

The second fiber laser in the setup has two components: a narrow-band, low-power source, followed by a high-power erbium-doped fiber amplifier (EDFA). The narrow-band source is an NKT (formerly Koheras) E15 Adjustik laser that produces up to 65 mW at 1550 nm with a (specified) linewidth of < 160 kHz. Coarse (slow) tuning of this laser is performed with a temperature adjustment (range 298.8 GHz), and fine (fast) tuning can be performed with a piezo-electric transducer (standard-option tuning coefficient ~14 MHz/V, range 6.2 GHz). For the results presented here, the wavelength is set to 1549.850 nm (vacuum).

The EDFA is a Manlight amplifier, with an operating range of 1545 nm to 1565 nm, and an output power of up to 4.57 W. The maximum input power to the EDFA is 32 mW, so that in principal we could drive two such EDFA's with a single Adjustik laser. The ratio of signal to ASE increases as the output power is increased up to a specified value of 23.8 dB at maximum output. The EDFA's output port has an angled-FC connector to which a high-power fiber patch cable is attached. The light exiting this fiber (NA = 0.14) is collimated with a

---

[1] A range of commercial equipment, instruments, materials, and suppliers are identified in this paper to foster understanding. Such identification does not imply recommendation or endorsement by the National Institute of Standards and Technology, nor does it imply that the materials or equipment identified are necessarily the best available for the purpose.

Thorlabs collimator package (model F260APC-1550, $f$ = 15.58 mm, NA = 0.16). The resulting collimated beam size (radius) is equal to 1.18 ± 0.02 mm. For optimum SFG, the waist size of the 1550 nm beam in the PPLN crystal is adjusted with a telescope. The output beam-profile quality parameter $M^2$ for both fiber lasers is less than 1.05.

The output fibers on both lasers are not polarization-maintaining, so a warm-up period of approximately one hour is required to stabilize polarizations. The polarizations are then adjusted with pairs of zero-order wave-plates to achieve maximum transmission through optical isolators (OFR, isolation > 33 dB) that protect the lasers from back reflections.

The two beams are overlapped on a dichroic mirror (from CVI Melles Griot) and focused into a periodically poled lithium niobate (PPLN) crystal, custom produced by Stratophase (now by Covesion). The PPLN crystal has dimensions ($\ell \times w \times h$) 40 × 10 × 0.5 $mm^3$, and three poled channels with periods of 10.90 µm, 10.95 µm and 11.00 µm, each 1 mm wide, that run the length of the crystal. The input and output facets are anti-reflection coated (reflectivity < 1 %) for 1550 nm, 1051 nm and 626 nm. The crystal is mounted with its large-area surface parallel to the optical table, and in this orientation, optimum sum-frequency generation is achieved with vertical polarization for both pump and signal beams. For the results presented here, we use the 10.90 µm channel and heat the crystal to 196.5 °C with a Covesion oven (model PV40) and a Thorlabs heater controller (model TC200). A benefit of operating at this (relatively high) temperature is that photorefractive damage caused by the visible light is not observed [15]. The FWHM of the temperature tuning of the quasi-phase matching is approximately 0.5 °C; thus the 0.1 °C set-point resolution of the heater controller is adequate, but not ideal.

Based on the Boyd and Kleinman model [15], optimum SFG occurs when the confocal parameters $b$ (equal to twice the Rayleigh length) for the pump and signal beams are equal, and when the focusing parameter is $\xi = \ell/b = 2.84$. Fortunately, near the optimum focusing parameter, the SFG conversion efficiency is a slowly varying function of this parameter, so that achieving good focusing is relatively straightforward. To optimize the SFG, we start with comparable waist sizes for the pump and signal beams. In our setup, the 1051 nm (pump) beam is focused by a $f$ = 12.5 cm lens. In free-space this lens produces a 40 ± 3 µm waist (measured before the PPLN crystal is installed). Optimum conversion is observed when the lens is positioned a distance 11.5 ± 0.1 cm from the front surface of the PPLN crystal. We calculate that this corresponds to a 58 ± 5 µm waist within the crystal, positioned midway along the crystal's length. A Galilean telescope in the 1550 nm (signal) beam path enables waist-size adjustment for SFG optimization (see Fig. 1). This consists of a –5.0 cm plano-concave lens and a 6.0 cm plano-convex lens. After optimization, the 1550 nm beam waist in free-space is 45 ± 3 µm (measured by deflecting the beam), corresponding to 66 ± 5 µm waist within the crystal. This is achieved with a 4.0 ± 0.2 cm spacing between the two telescope lenses, and a distance from the converging lens to the front surface of the PPLN crystal of 171 ± 2 mm. The waist sizes are consistent with the Boyd & Kleinman prediction that optimum conversion occurs when the Rayleigh length of the pump and signal beams are equal. The focusing lenses are all mounted on translation stages so that the waist positions can be optimized.

A plot of SFG output power versus the product of pump and signal input powers is shown in Fig. 2. From a straight-line fit to the plot we determine the SFG efficiency

$$\eta = \frac{P_{626}}{P_{1051} P_{1550} \ell} = (2.7 \pm 0.1)\% \text{ W}^{-1}\text{cm}^{-1}, \quad (1)$$

where $P$ refers to the power in beams at each of the three wavelengths. For our wavelengths and focusing parameter $\xi = 0.9 \pm 0.2$, the Boyd and Kleinman model predicts a conversion efficiency that is approximately 30 % larger than what we measure. This difference, between our measured and predicted efficiency, is consistent with measurements and corresponding calculations reported elsewhere, for example [13, 16-17]. Possible reasons for the discrepancy include diffraction loss at the edges of the PPLN crystal, an uneven temperature profile within the crystal, uncertainty in the value of the effective nonlinear coefficient, imperfections in the periodic poling, and/or non-ideal beam overlap and sizes. With our fiber lasers both running at maximum power, and taking into account ~10 % estimated losses in both the pump and signal beam before they enter the crystal (nearly all in the optical isolators), we achieve 24 % conversion, generating 2.0 W of output at 626 nm with a total of 8.5 W NIR input. For this measurement, mirrors with high reflectivity at 626 nm and high transmission at 1051 nm and 1550 nm (supplied by Laseroptik) were used to separate the visible light from transmitted NIR pump and signal light.

When the 2 W SFG output power was monitored over several days, we did not observe the sudden power decreases reported by Bosenberg *et al*. with 2.5 W output at 629 nm [18]. Any drift in power that we observe (typically < 5 %) is due to slow changes in polarization from the non-polarization-maintaining fibers. These changes are associated with small laboratory temperature variations (< 1 $^{o}$C), and an adjustment of the wave-plates after the optical fibers restores the output power. This minor problem could be improved with polarization-maintaining fibers (which a number of fiber laser manufacturers can now supply).

With the previously specified fiber laser wavelengths, the sum-frequency output wavelength is 626.342 nm (vacuum). After doubling, this corresponds to a detuning of +80 GHz from the $^{9}$Be$^{+}$ $2s^2S_{1/2}$ to $2p^2P_{1/2}$ transition, which is typically what we have used for stimulated Raman transitions. However, the tuning ranges of our fiber lasers allow SFG from 626.119 nm to 626.445 nm. As illustrated in Fig. 3, this means that after frequency-doubling (see below) we can tune to the $^{9}$Be$^{+}$ Doppler cooling transition, to the repumping transition, or to any of a range of detunings for stimulated Raman transitions.

## 3  Cavity-enhanced, second-harmonic generation of 313 nm light

For SHG into the UV there exists a range of nonlinear optical (NLO) materials [19], but the most readily available options are LBO (LiB$_3$O$_5$) and BBO (β-BaB$_2$O$_2$). Both these materials have high laser-induced damage thresholds. LBO has good transparency down to 160 nm, but can be phase-matched only to as low as approximately 275 nm. BBO has good transparency down to only 190 nm, but can be phase-matched to approximately 205 nm. LBO has a smaller birefringent walk-off angle, but BBO has an effective NLO coefficient that is approximately twice that of LBO (and SHG efficiency is a quadratic function of this parameter).

In addition, in earlier testing of LBO we observed long-term surface degradation. Another possible NLO material is KDP, which can be phase-matched down to approximately 260 nm, but its effective NLO coefficient is approximately six times smaller than that of BBO. KTP on the other hand has a slightly higher NLO coefficient than that of BBO, but it is not transparent below approximately 350 nm.

In the setup described here, the 626 nm output of the SFG setup is frequency-doubled to 313 nm by use of a Brewster-angled BBO crystal within a ring cavity. This enhancement cavity is resonant at the pump wavelength and essentially transparent at the second-harmonic wavelength. We opted for a Brewster-cut crystal, rather than an anti-reflection (AR) coated square-cut crystal, due to concerns about possible damage to an AR coating at high UV power. The BBO crystal (provided by Castech) has dimensions ($\ell \times w \times h$) 10 × 4 × 4 mm$^3$ and is cut for critical (or angle) type I phase-matching at room temperature [20]. In this configuration, two photons at the fundamental wavelength, both polarized normal to the crystal's principal plane (the plane formed by the crystal's optic axis and propagation vector), generate a single photon polarized parallel to the principal plane. This arrangement is well suited to the situation in which the crystal for SHG has input and output surfaces at Brewster's angle, and the intensity of the fundamental beam is enhanced by a resonant cavity. There are, however, two disadvantages. First, although there is minimal reflection loss of fundamental light from the Brewster-angled crystal surface, the second harmonic is polarized orthogonal to the fundamental, so that in our setup approximately 16 % is reflected from the crystal surface. Second, as discussed below, birefringent walk-off limits the SHG efficiency.

Using the Sellmeier equations [21-22] to determine the ordinary and extraordinary indices of refraction for BBO, we calculate the phase-matching angle $\theta$ (the angle between the optic axis and the propagation vector) for the wavelengths of interest, and Brewster's angle for the fundamental wavelength. These parameters determine the required crystal cut: Brewster's angle = 59.1$^{o}$, $\theta$ = 38.4$^{o}$ and azimuthal angle $\varphi$ = 0$^{o}$. Following the notation of Boyd & Kleinman, the birefringent walk-off angle is $\rho$ = 80 mrad, and the related crystal parameter is $B \equiv \frac{1}{2}\rho(k_1\ell)^{1/2} = 16.4$ (compared with $B = 0$ for no birefringence), where $k_1$ is propagation vector for the fundamental within the crystal. As expected for BBO, the walk-off is relatively large.

The Boyd & Kleinman model for SHG optimization considers a Gaussian beam focused into a nonlinear uniaxial crystal, and in the absence of birefringent walk-off the model predicts optimum SHG for the same focusing parameter as for parametric generation i.e., $\xi = 2.84$. In the case of large walk-off (such as ours), relaxed focusing is necessary to maintain phase matching, and the optimum focusing parameter reduces to an asymptotic value at large $B$ of $\xi = 1.39$. After Boyd and Kleinman's pioneering work, it was discovered that use of an elliptical beam profile, in which the focusing is tighter in the transverse direction normal to the principal plane, can improve conversion efficiency [23-26]. More recently, Freegarde *et al.* presented a generalized model for optimizing SHG with elliptical focusing [27], and characterize their focused elliptical beam by two transverse focusing parameters $\xi_x$ and $\xi_y$. They predict up to 30 % conversion enhancement with an elliptical beam.

Our enhancement cavity design adopts the approach of Freegarde *et al.* to optimization. Ideally, we want to alleviate the walk-off problem with a cavity configuration that provides the astigmatic focusing calculated by Freegarde *et al.* In addition, we want to compensate for the astigmatism introduced by the intra-cavity Brewster-angled crystal. As pointed out by Freegarde *et al.*, for negative uniaxial crystals (such as BBO), this ellipticity is unfortunately oriented orthogonal to the direction that improves conversion efficiency. In practice, with a cavity based on off-axis spherical mirrors, there is a limit to the amount of ellipticity that the cavity can induce before becoming unstable. Effectively, the off-axis mirrors have different focusing in, and normal to, the plane of incidence, so that two stability conditions must be satisfied.

A schematic of the enhancement cavity is shown in Fig. 1. It is a bow-tie configuration with two off-axis spherical mirrors and two plane mirrors. For ease of alignment and robust operation, the design has the astigmatic waist inside the crystal to provide maximum immunity to small changes in the spacing between the two curved mirrors. We begin the design process by determining the off-axis angle and mirror separations that produce a Boyd & Kleinman optimum (circular) waist within the crystal. These parameters are then the starting point for a search to find new parameters that produce an optimum elliptical waist. A key advantage of requiring the stability regions to maximally overlap is that the secondary waist within the cavity is almost circular, so that mode-matching into the cavity can be done in the usual way with a pair of spherical lenses, as shown in Fig. 1.

The calculated optimum cavity dimensions are as follows. The distance between the spherical mirrors and the crystal faces is 24.2 mm. The long path length, between the spherical mirrors (including the reflections from the two plane mirrors) is 527.6 mm. The full off-axis angle $\alpha$ (see Fig. 1) for the spherical mirrors is 30.0°. In this configuration, the horizontal and vertical waists of the 626 nm beam within the crystal are respectively 26.0 µm and 16.8 µm, and the secondary waist (positioned midway between the plane mirrors) has sizes in the horizontal and vertical directions of respectively 218.6 µm and 217.8 µm. To minimize the horizontal space occupied by the doubling cavity, we numerically investigated the effect of reducing the long path length and discovered that the conversion efficiency is relatively insensitive to this parameter. Therefore, in our final design we reduced the long path length to 290 mm, so that the footprint of the enhancement cavity is 16 × 22 cm. We predict that this reduces the conversion efficiency by 8 %. Repeating the optimization with this shorter path length, the distance between the spherical mirrors and the crystal faces remains unchanged at 24.2 mm, but the spherical mirror full off-axis angle is $\alpha = 28.6°$. The horizontal and vertical waists of the 626 nm beam within the crystal become 36.7 µm and 23.6 µm (respectively), and the secondary waist sizes in the horizontal and vertical directions become 155.3 µm and 154.3 µm (respectively). Overall, the ellipticity within the crystal is essentially unchanged (e = 1.55) by the reduction in path length, but the waist sizes are larger, leading to the modest drop in conversion efficiency.

The bow-tie beam path lies in the horizontal plane and the input light at 626 nm is horizontally polarized. To reduce output beam divergence, spherical mirror M1 is meniscus, and for convenience M2 identical to M1. The front

(reflective) surfaces of mirrors M1 and M2 are concave with a radius of curvature of -50 mm, and the back surfaces are convex with radius of curvature of 50 mm. The diameters and thicknesses of these mirrors, as well as the plane input-coupling mirror (M3), are 12.7 mm and 6.25 mm, respectively. The small plane mirror (M4) has a diameter of 6.0 mm and a thickness of 2.0 mm, and it is mounted on a piezoelectric transducer (PZT) from Thorlabs (part #: AE0505D08F). The transmission of the input-coupler mirror was chosen to match the estimated total loss of all other elements in the cavity, including that due to the SHG of a single pass through the crystal at 1 W of incident power. This "impedance-matching" maximizes the incident light that is coupled into the cavity. The input-coupling mirror has a transmission of 1.6 % at 626 nm, and the other mirrors are high-reflectors (reflectivity $R > 99.9$ %) at 626 nm. All the mirrors have good transmission at 313 nm ($T = 95$ %). The mirror substrates are provided by Laseroptik, and the coatings are done by ATFilms. In testing of stock mirrors for input coupling, it was discovered that it is necessary to use a mirror with a hard coating (e.g., ion beam sputtered) to avoid deterioration due to UV damage at high powers. To correct the astigmatism in the UV beam exiting the enhancement cavity, a $f = 7.5$ cm cylindrical lens is positioned 5.8 cm from the output mirror's (M1) front spherical surface.

The mirrors are installed into Lees mirror mounts (supplied by Linos/Qioptiq), and these are anchored to a 12.7 mm thick CNC-machined aluminum-alloy baseplate. Dowel pins fix the mirror mounts into the correct positions. The BBO crystal is mounted on a New Focus 5-axis aligner (model 9081). Since BBO is hygroscopic, and its surfaces become "fogged" in humid air, the crystal is housed in a partially-sealed enclosure that is purged with a gentle flow of dry oxygen [28]. As mention above, there is significant reflection of the UV light at the crystal surface (see the dashed arrow in Fig. 1), and a hole in the side of the BBO mount gives us access to this secondary output beam. The cavity is servo-locked on resonance with the input light by use of the Hänsch-Couillaud method [29], with proportional-integral feedback to the small PZT-mounted mirror (M4). The bandwidth of the servo-control loop is 50 kHz. If the servo controller becomes unlocked from mechanical perturbation, a computer control system [30] zeros the integrator, finds the cavity peak, and relocks the servo. To dampen mechanical resonances, the PZT element is fitted into a lead cylinder that is in turn fitted into the mirror mount.

Adams and Ferguson present a simple model for cavity-enhanced SHG [31]. For low incident power, in which the intra-cavity loss due to SHG is small, the circulating power increases linearly with the incident power, so that the second-harmonic power increases quadratically. However, for higher incident power, the loss due to SHG becomes significant, the intra-cavity power is strongly attenuated by the conversion process, and second-harmonic power increases almost linearly with incident power. A plot of output power at 313 nm versus input power at 626 nm is shown in Fig. 4. As expected for our parameters, the power of the second harmonic is almost linear with the incident power. The maximum output beam power we obtain is 760 mW, and the linear region of the curve corresponds to a conversion efficiency of 42 %. If we include the secondary output beam reflected from the crystal surface, the conversion efficiency is 50 %. At the highest incident power, the conversion efficiency decreases. This behavior is likely due to thermal lensing in the BBO crystal, which compromises the mode-matching into the enhancement cavity [32]. Even at the highest powers, the peak intensities of the 626 nm and 313 nm light within the crystal are both many orders of magnitude below the damage threshold for BBO reported by suppliers and in the review article by Nikogosyan [33]. The output intensity is stable within 4%, and with intensity stabilization via an EOM and servo control with a bandwidth of 300 kHz, this is reduced to approximately 1 %.

Finally, the laser system has been used to perform Raman sideband transitions and to cool $^9$Be$^+$ ions to the motional ground state of a harmonic trapping potential [34].

### 4  Conclusion

In summary, we have developed a solid-state laser system for generating high power at 313 nm – the wavelength needed for laser cooling and manipulation of trapped $^9$Be$^+$ ions. SFG with near-infrared pump and signal sources is used to produce 626 nm, and cavity-enhanced SHG converts this visible light to 313 nm. Commercial fiber lasers rated for 5 W at 1550 nm and 1051 nm are frequency-summed in PPLN to produce 2 W at 626 nm, corresponding to a conversion efficiency of 24 %. SHG to the UV is then implemented with a Brewster-angled BBO crystal in a resonant enhancement cavity. The SHG conversion efficiency is 42 % for a wide range of incident powers, and up to 750 mW at 313 nm is produced. This output is five times more power than what is predicted to be necessary for fault-tolerant quantum gate operations based on stimulated Raman transitions with trapped beryllium ions.

**Acknowledgements** This work was supported by IARPA, DARPA, and the NIST Quantum Information Program. We thank members of the NIST Ion Storage Group for helpful suggestions and advice, especially Till Rosenband, Ulrich Warring, Yves Colombe and Robert Jordens. We thank Christian Rahlff at Covesion and Hsiang-yu Lo at ETH Zurich for helpful discussions. This paper, a submission of NIST, is not subject to US copyright.

**FIGURE 1.** Schematic diagram of the optical setup. Near-infrared light from two fiber lasers is frequency summed in PPLN to produce red (626 nm) light, and this is then frequency doubled in BBO to the UV (313 nm). The 4-mirror bow-tie-shaped enhancement cavity is frequency locked to the 626 nm input light. SMF – Single mode fiber, COL – fiber collimator, λ/4 – quarter-wave plate, λ/2 – half-wave plate, ISO – optical isolator, DM – dichroic mirror, PPLN – periodically-poled lithium niobate crystal (40 mm long), M1 & M2 – meniscus mirrors with -50 mm radii of curvature on the reflective (front) surface and +50 mm on the back surface, M3 – plane input coupler mirror (T = 1.6 %), PZT – piezoelectric transducer, M4 – plane PZT-mounted "tweeter" mirror for cavity lock, BBO – Brewster-angled β-BaB$_2$O$_4$ nonlinear crystal (10 mm long), dashed arrow from BBO crystal – 313 nm light reflected from the crystal surface (R = 16 %), $f_{CYL}$ – focal length of cylindrical lens, WP – Wollaston prism, PD – photodiode. Hänsch-Couillaud cavity lock circuitry includes difference (–) and proportional-integral (PI) feedback, and high-voltage amplification (HV).

**FIGURE 2.** 626 nm sum-frequency output power versus the product of input pump and signal powers. For these data, the 1550 nm beam power was fixed and the 1051 nm beam power varied. However, the same efficiency was measured when the 1550 nm beam power was fixed and the 1051 nm beam power varied. The horizontal error bars include the uncertainty associated with polarization changes in the output of the fiber lasers, as well as the uncertainty in the near-infrared laser power measurements. The vertical error bars include the uncertainty associated with residual PPLN crystal temperature changes, as well as the uncertainty in the sum-frequency measurement of output power.

**FIGURE 3.** Energy level diagram for $^9$Be$^+$, illustrating the tuning range of the laser system with respect to the D$_1$ and D$_2$ transitions.

**FIGURE 4.** 313 nm second harmonic output power versus 626 nm input power. The dominant contributor to the error bars is the uncertainty in the laser power measurements.

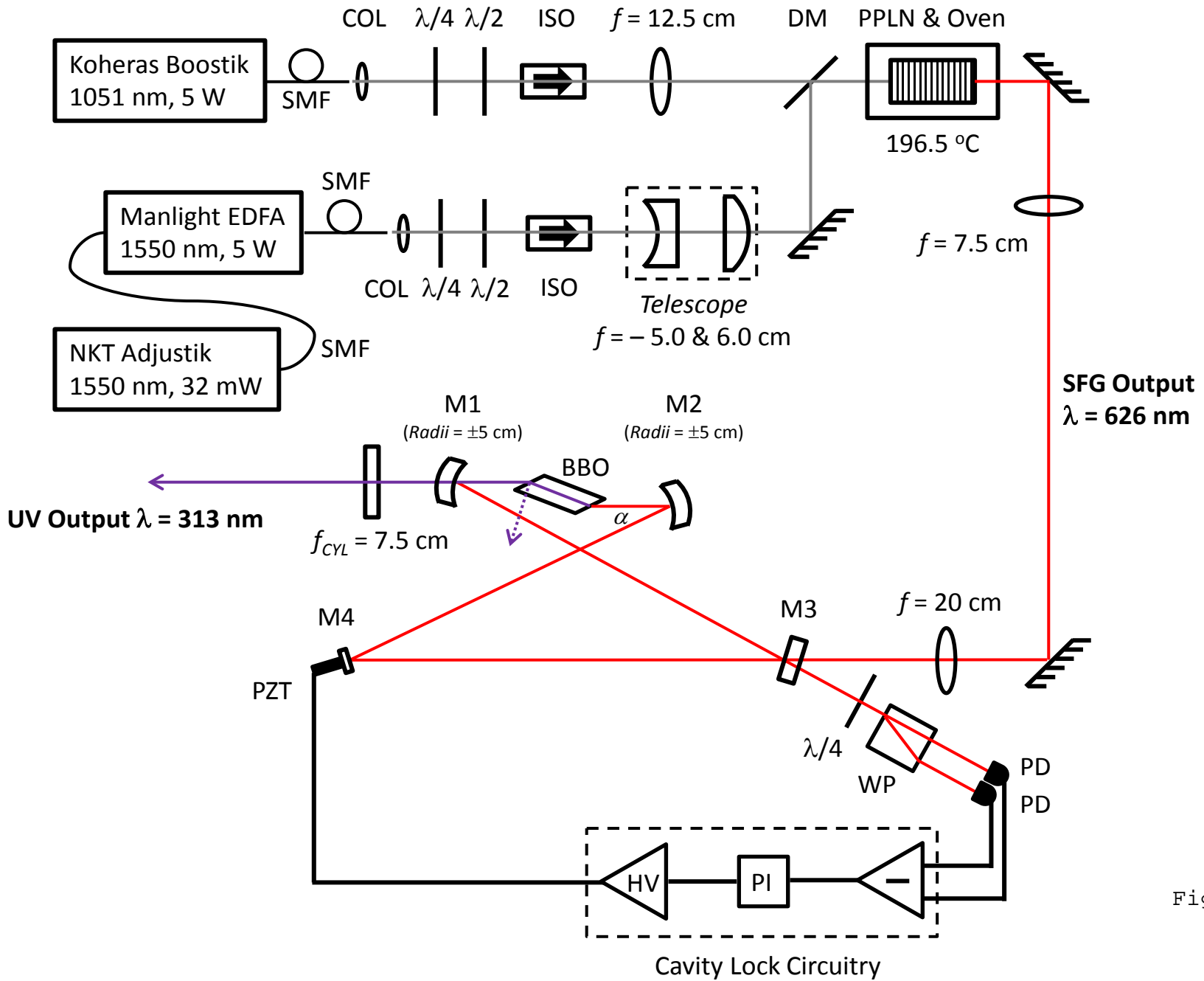

Fig. 1

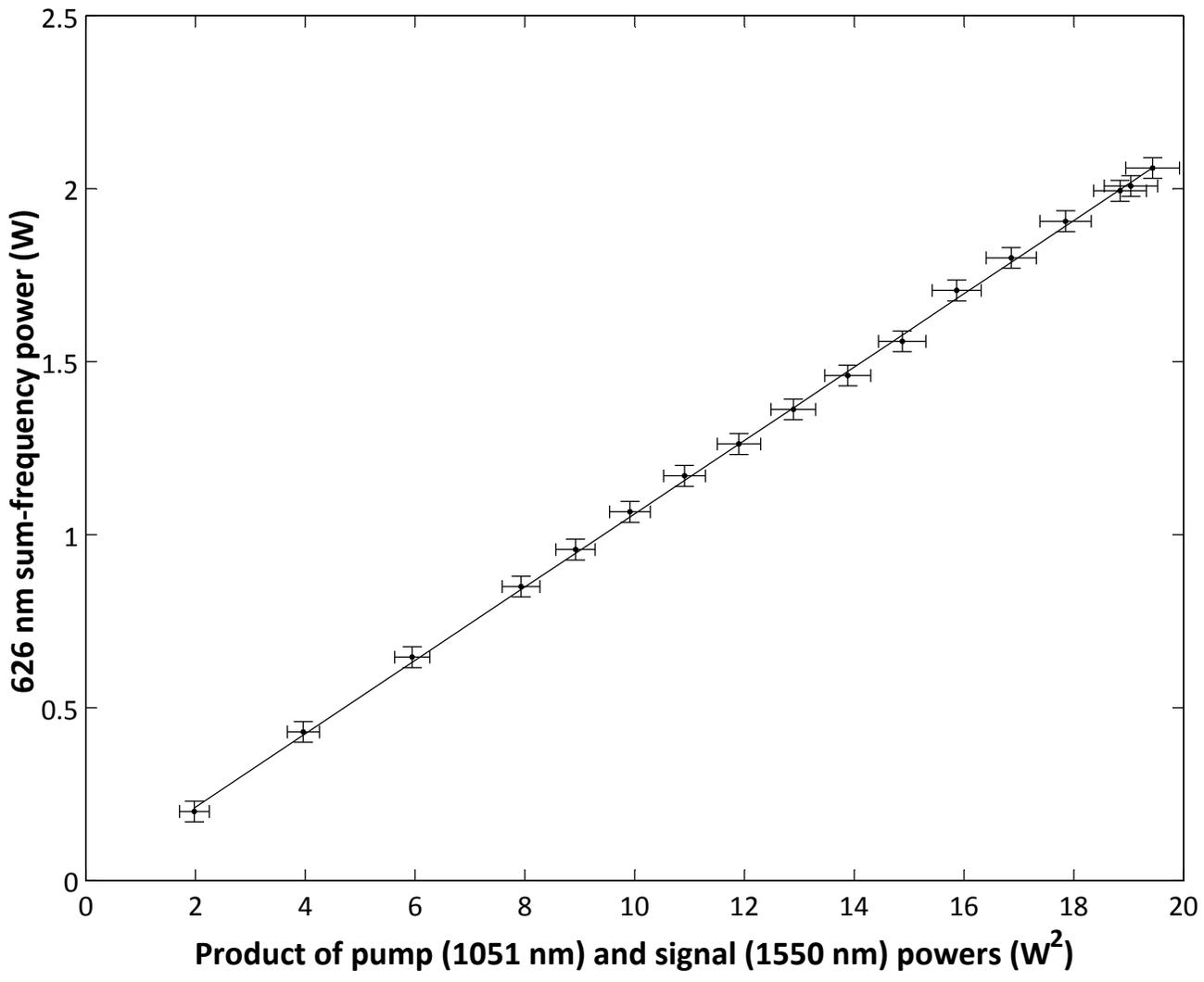

Fig. 2

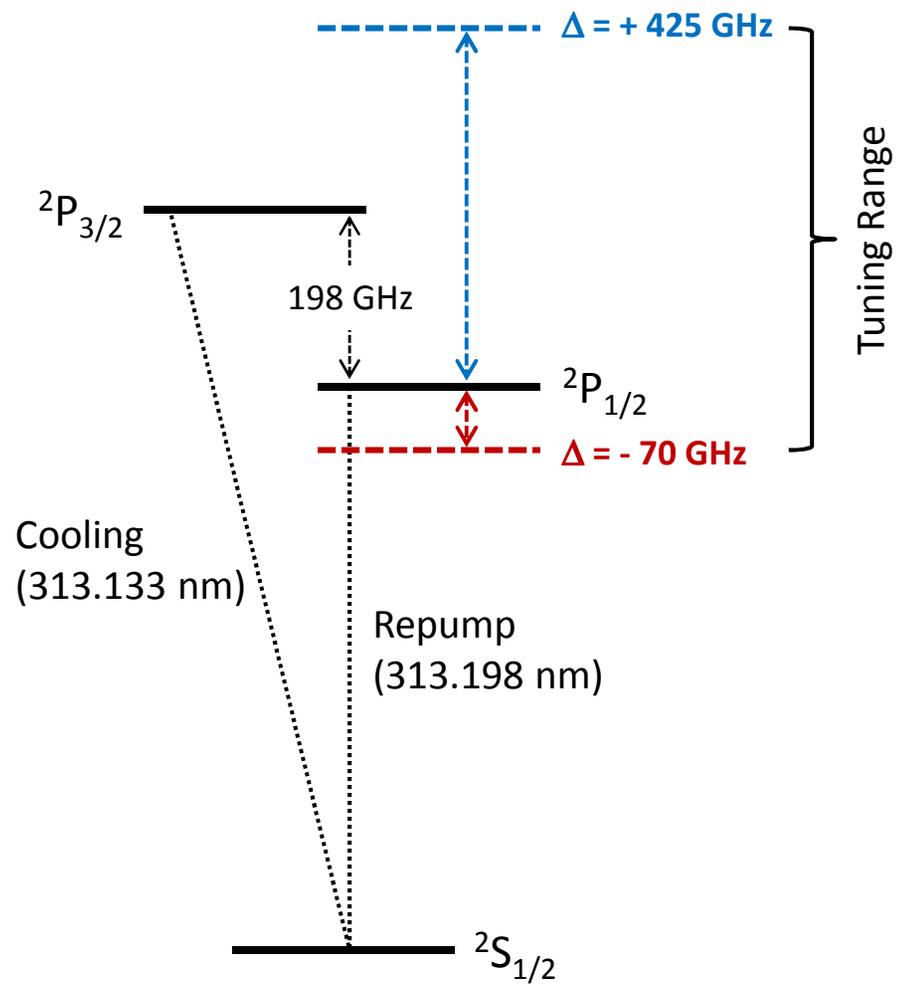

Fig. 3

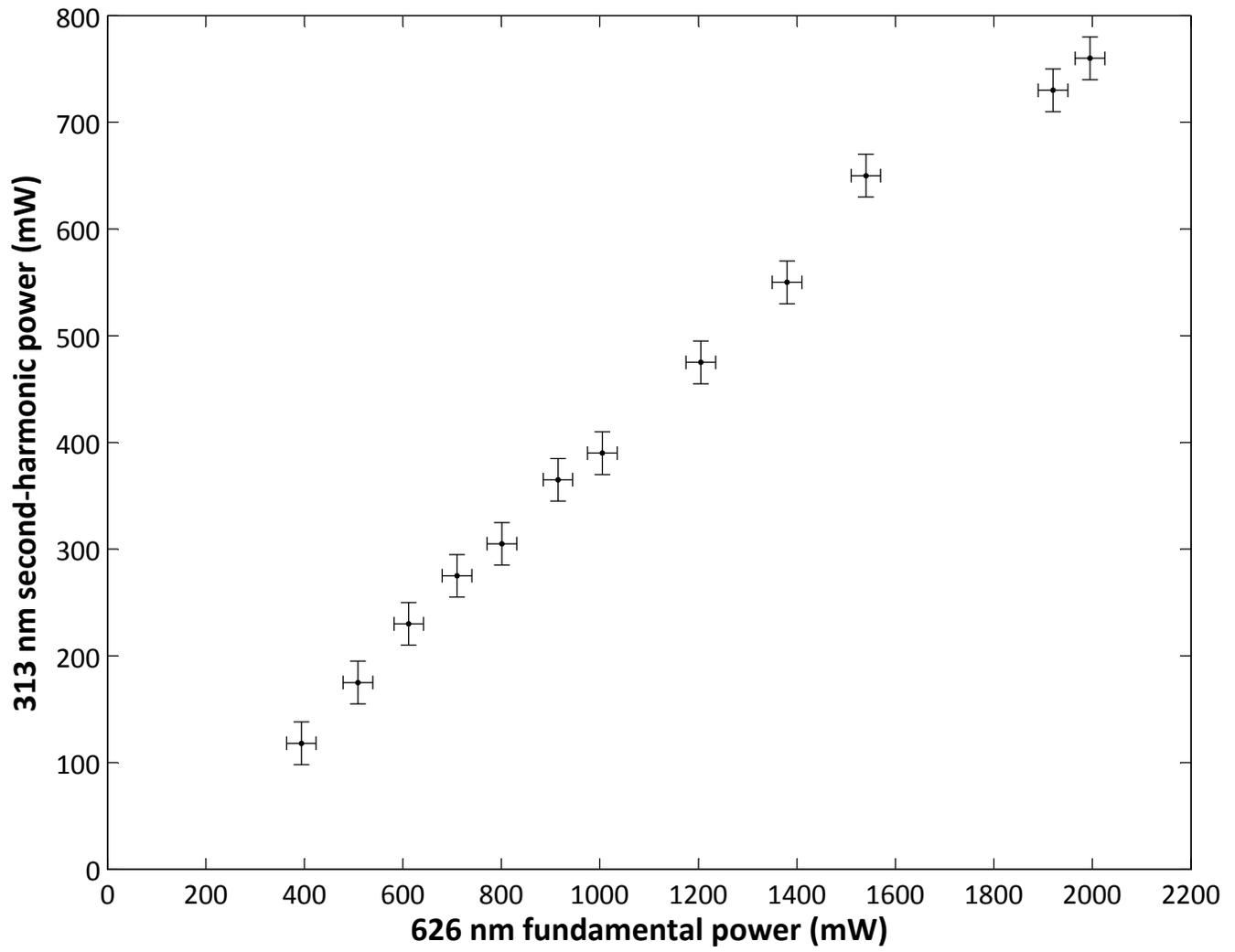

Fig. 4